# TOWARDS A VIRTUAL DATA CENTRE FOR CLASSICS

## TOBIAS BLANKE, MARK HEDGES, (AND SHRIJA RAJBHANDAR)

*1. Introduction*

For many years Classics researchers have been producing a variety of digital outputs, whether in the form of relational databases, corpora of texts marked up in XML (Extensible Markup Language), or other formats. Naturally, these resources tend to focus on specific research topics that reflect the interests of their creators, whether in terms of the nature of the source material, or the time periods, communities, and geographical areas addressed; nevertheless, they are reusable resources that support research beyond that intended, or even envisaged, by their creators. Moreover, the content of these various digital resources is often conceptually related, each representing a small part of the increasingly rich data landscape available for the ancient world, and they would be of much greater utility to researchers if they could be linked up in a way that allowed them to be explored as a unity. However, while the resources may be reusable, the variety of data representations and formats used militates against such an integrated view. This is the challenge of interoperability – without interoperability, each resource remains an island, which can be combined with other resources only with a great deal of effort on the part of the researcher.

One way of approaching the interoperability challenge has been standardization. Many discussions have taken place (and not just in the humanities) about establishing standards for the creation of digital resources, with the aim (among others) of facilitating the creation of highly interlinked corpora. An important example of this from Classics is EpiDoc,[1] which provides standards and guidelines for the mark-up and interchange of inscriptions and other ancient documents which conforms to the TEI (Text Encoding Initiative) XML guidelines.

However, although the development of standards such as EpiDoc is an important step forward, standardization is unlikely to solve all issues around linking up heterogeneous data in the humanities, for a number of reasons. Firstly, there exists already a great deal of legacy data in diverse, non-standard, and often obsolete formats.[2] Secondly, users have first to be trained in the correct application of a standard, which requires potentially a large investment in terms of time and money that not all projects may be able to accommodate. Thirdly, even when standards are used, the sheer variety of the data means that there is significant flexibility in how the standards are applied (a selection of TEI documents, for example).

---

[1] EpiDoc website: <http://epidoc.sourceforge.net>.

[2] T. Blanke, M. Hedges, and S. Dunn, 'Arts and humanities e-science: current practices and future challenges', *Future Generation Computer Systems* 25. 4 (2009) 474-80.

Finally, standards are generally developed within particular disciplines or domains, such as (in the example of EpiDoc) epigraphy, whereas research is often inter-disciplinary, making use of varied source material and incorporating data conforming to different standards. There will inevitably be diversity of representation when information is gathered together from different domains and for different purposes, and consequently there will always be a need to integrate this diversity.

The approaches that we describe are based on the principle of respecting the integrity of existing representations of data, while virtualizing data and services over the web – that is to say, creating a virtual version of a resource by means of an abstraction layer that is independent of the underlying data structures and storage systems, allowing heterogeneous resources to be treated in a common fashion. It is outside the scope of this work to address the very broad range of digital resources available in Classics; rather, we investigated the issues raised by integrating structured datasets relating to ancient documents, albeit structured datasets that contained a significant quantity of unstructured text and structured data. Specifically, we used datasets relating to epigraphy and papyrology, although the issues raised are of relevance to other datasets of analogous form. We present two case studies representing different approaches, the first using data-grid technologies to provide integrated views of resources, the second enabling integrated content-based retrieval of resources. We consider the two approaches to be complementary, each providing in different ways dynamic integrated views over the data, and reducing uncertainties about the information by linking it to related information in other sources. We elaborate on this further elsewhere.[3]

The structure of the chapter is as follows. First, we present the background to the work, in Section 2 describing the specific collections used for our experiments and examining the interoperability issues that they raise, and in Section 3 introducing the technologies and approaches that we used to virtualize the collections. These approaches are addressed in two case studies presented in Sections 4 and 5. Finally, in Section 6, we deliver the vision of a Classics virtual data centre that emerged from our investigations. In Section 6, we also discuss in which related disciplines like archaeology such a service already exists. We suggest a virtual data centre for Classics, as the data resources are smaller and more spread out. Also, a virtual data centre could be more easily embedded in larger initiatives and would be therefore cheaper to maintain. Finally, all the experiments presented here will be integrated into the architecture discussion and realization of the emerging European infrastructure for digital arts and humanities, *DARIAH*.[4]

*2. The data and its challenges*

The digital resources used in our experiments primarily concern ancient documents, in particular classical epigraphy and papyrology, and in terms of formats included relational databases with different schemas and implemented using different data technologies, as well as a corpus of XML data. Specifically, the resources used were:

---

[3] T. Blanke and M. Hedges, 'Humanities e-science: from systematic investigations to institutional infrastructures', in *E-SCIENCE 10: Proceedings of the Sixth IEEE International Conference on e-Science* ed. by? (Washington, DC 2010) Pages?.

[4] *DARIAH* website: <http://www.dariah.eu>.

- The *Heidelberg Gesamtverzeichnis der griechischen Papyrusurkunden Ägyptens* (*HGV*),[5] a Filemaker Pro database containing metadata on some 55,000 papyri, mostly from Roman Egypt and its environs, including bibliography, dates, and places (*e.g.* findspots and provenances);
- *Projet Volterra*,[6] a Microsoft Access database of Roman legal pronouncements and associated metadata, from various sources, whether epigraphic, papyrological, juristic, or literary;
- The *Inscriptions of Aphrodisias* (*IAph*),[7] an XML corpus of 1500 inscriptions from the ancient city of Aphrodisias in Asia Minor, including transcribed texts and metadata marked up using EpiDoc TEI, as well as images.

These three datasets vary significantly, both in terms of their information content and the formats used for implementing them; however, there is sufficient overlap in content to make integrating them profitable to researchers. For instance, although the *Volterra* collection is specific to legal texts, it contains some papyri and therefore some references to find-spots that also occur in the *HGV* metadata. Similarly, although none of the *Volterra* texts are inscriptions from Aphrodisias, there are attestations of persons that appear in both the *Volterra* and *IAph* texts (especially in the late antique period, for which the Aphrodisias material is most richly annotated). The *IAph* and *HGV* collections do not have any content in common, but the categories that are used to organize the texts within these resources have a certain overlap, for example letters, decrees, honours, contracts. All three datasets overlap fairly closely in date, and have similar (but not identical) mechanisms for recording dates, date ranges, periods, and uncertainty of dating.[8] Cross-corpus searches based on these areas of overlap should provide realistic tests of our approaches, as well as yielding potentially useful scholarly results. Note, however, that these three resources are just three examples from a much larger pool of related datasets that might have been included in a larger-scale integration project, and were selected in order to investigate the feasibility of our approaches and to identify issues that might arise.

We may make the following observations about these three datasets and the researcher's broader data environment that have consequences for data interoperability:

- Data formats are very diverse, and involve multiple media and standards;
- Databases rarely follow standard database schemas. The use of mark-up can vary significantly, particularly in resources developed before much effort had been made towards standards (such as EpiDoc), but natural variation occurs even in applying these standards;
- The material may be highly complex, with many structural and semantic relationships both internal – for example within a TEI document – and contextual. The interpretation of an object (*e.g.* an inscription) may depend on its relationships to other resources and collections (*e.g.* other inscriptions, literary texts, archaeological surveys, concordances), which are moreover not necessarily digital;

---

[5] *HGV* website: < http://www.rzuser.uni-heidelberg.de/~gv0/>.

[6] *Projet Volterra* website: <http://www.ucl.ac.uk/history2/volterra>.

[7] *Inscriptions of Aphrodisias* website: <http://insaph.kcl.ac.uk>.

[8] A particularly challenging issue being investigated is that of handling different levels of uncertainty in temporal data: some dates are extremely precise – even to the day – whereas many others are very vague – perhaps to a span of 50 or 100 years.

- Data may be incomplete, indeed incompletable – the capture of the data cannot be repeated nor the data enhanced to fill in the gaps. For example, an inscription may be damaged, a papyrus's provenance not recorded, a corpus of texts fragmentary, the date of an event unknown;

- Data may be fuzzy or uncertain, or even contradictory. For example, there may be several sources for the date of an event, with various degrees of precision (to the year, to the decade) and various degrees of reliability;

- The resources are not easily available for use. In many cases, they are locked away on departmental machines; in other cases they are 'published' on a web site but not in a way that makes the resources particularly usable by a researcher;

- Even when a resource is available it is often available only in isolation. Many of these resources may be regarded as fragments of a larger picture, and would have vastly more value if researchers could have access to this larger picture, rather than just the parts;

- The resources may be owned by different communities and subject to different rights; the scholars who created them may be unwilling to accept anything that affects the integrity of the original resources. Consequently, any integration initiative must respect this autonomy and integrity, if it is to be successful;

We would not argue that such issues arise with respect to data only in Classics, nor that all Classics data can be characterized in this way. These issues will, however, be recognized by a significant number of researchers in many humanities disciplines.[9]

*3. Virtualization approaches*

Our general aim is to enable sharing of heterogeneous data resources in Classics in an integrated fashion, rather than as a number of isolated resources. Our broad approach may be described as one of virtualization of data and of access to data, by means of abstracted and standardized interfaces and protocols. Virtualization describes in computing all approaches that create a virtual version of a physical resource and goes back to the early days of computing with hardware virtualization strategies. In our case, instead of the actual data resources, the users can directly interact with virtual combination of them. Data can generally be *virtualized* in relation to several aspects:

- Location. Access is provided independently of where the datasets reside.
- Autonomy. Data may be governed by independent management regimes, owned by different communities and subject to different rights. Access is made more uniform while respecting the integrity of the original data and the environments in which it is managed, so that access to the data is in accordance with the terms of the data holders.

---

[9] For an extensive discussion and systematic analysis of the characteristics of humanities data see also M. Doerr, 'The CIDOC conceptual reference module: an ontological approach to semantic interoperability of metadata', *AI Magazine* 24.3 (2003) Pages.

- Heterogeneity, both the infrastructural heterogeneity of the storage, and the structural heterogeneity of the data. Virtualization means that datasets do not need to be accessed in possibly idiosyncratic ways.

Although all three are relevant for our work, the third aspect is the key one for integrating diverse resources. Virtualization can hide 'irrelevant' (for whatever purpose we have in mind) differences between data resources, giving the user more seamless access to them. Distributed, autonomous, and heterogeneous datasets can be federated and regarded as a single resource, enhancing the visibility of the data and multiplying the uses to which it can be put. Virtualization offers, therefore, new ways of defining interfaces between datasets, where irrelevant aspects are ignored and the common information across the datasets retained. In this paper we discuss two approaches to virtualizing the data resources. The first publishes the resources as data services that expose datasets in a standard, relational, database-like way, while the second allows virtual representations of resources to be constructed by building common indexes from existing datasets.

*4. Virtualization case study 1: linking and querying ancient texts*

The *LaQuAT* (*Linking and Querying Ancient Texts*) project[10] investigated the use of the OGSA-DAI (Open Grid Service Architecture-Data Access and Integration)[11] middleware. OGSA-DAI is widely used for supporting virtual integration of diverse, distributed data resources, providing 'on-the-fly' common interfaces to data. Its primary focus was on relational databases, and it supports integrated views across many different database management systems, with a particular view to querying, transforming, and delivering data in different ways via a simple toolkit for developing client applications. It was used in the first instance to provide an integrated view across relational databases with different schemas, namely the *HGV* and *Projet Volterra* data resources. OGSA-DAI is designed to be extensible, and subsequently our work was extended to integrate the *IAph* XML corpus, providing an integrated view over the three structured data resources.

Figure 1 shows how the *LaQuAT* architecture integrates different database resources. OGSA-DQP (Distributed Query Processing) was the main abstraction mechanism, hiding the details of the database implementations from the user. Our approach to virtual data integration is thus to specify the local data sources as views over the global schema. Out of the separate databases we create a large, virtual one.

In the case of the integration of the *Volterra* and *HGV* data resources, two principal alternatives were discussed and evaluated. On the left hand side of figure 1, the architecture contains a single abstraction layer using OGSA-DQP, which hides the implementation details of the underlying databases. In addition, we bridge the language divide between the German *HGV* and the English *Volterra* data resources by using a *join* table to map between German and English keywords. On the right hand side, the architecture uses an additional OGSA-DQP abstraction layer to hide the fact that *HGV* is a German database. We decided that the former was the preferable solution, as access to the *join* table may be beneficial for other data resources.

---

[10] *LaQuAT* project website: <http://www.laquat.cerch.kcl.ac.uk/>. *LaQuAT* was funded by JISC (Joint Information Systems Committee).

[11] OGSA-DAI website: <http://www.ogsadai.org.uk>.

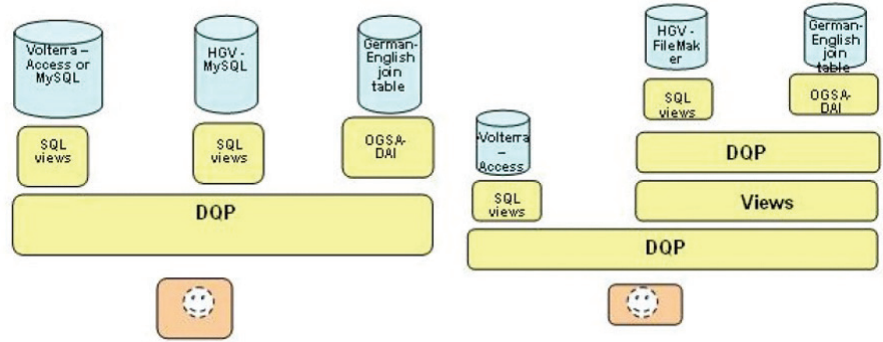

Figure 1: LaQuAT Architecture

OGSA-DAI uses SQL (Structured Query Language) views to hide the details of a data resource. In OGSA-DAI, everything from a standard database, to an XML file, to an indexed text resource will look to the user as if they were interacting with a single, large SQL data resource. To this end, OGSA-DAI generalizes the concept of an SQL view and virtualizes it. For *LaQuAT*, the following combination of traditional database technologies and OGSA-DAI technology will realize the virtualization of the data resources.

SQL views can handle the following requirements:

- Expose TEXT date column types as DATE date column types. In *Volterra*, for example, all date-related fields are defined as text fields in MS Access.

- Union ?unite? N tables so they are treated as a single table. This is standard-view functionality. However, some of the data resources like MS Access have very specific ways of realizing them.

- Expose German column and table names as English, handling any spaces and German characters.

OGSA-DAI DQP can additionally handle the following requirements:

- Expose multi-lingual column contents as English. This is done using the already-mentioned join table.

- Perform text searches over the contents of individual fields.

- Perform a join across databases.

*LaQuAT*'s results were promising from a Classicist's point of view, as new lines of enquiry by combining existing data resources could be explored, for example by discovering references to homonymous (and possibly identical) persons in different texts that could be dated to within a small number of years of one another. Nevertheless, *LaQuAT* also identified limitations to this approach to data integration in the case of humanities resources.

In general terms, OGSA-DAI is optimized for working with *data-centric* rather than *text-centric* resources. The distinction here is between resources that contain significant quantities of unstructured text (text-centric), and those that consist primarily of structured data such as numerical data, dates, or very short text fields. In the humanities, however, researchers work more commonly with text-centric resources, such as text documents, within which they want to find relevant information so that standard document retrieval techniques can be applied and adapted for dealing with the specifics of handling additional structural constraints.[12] Indeed, the limitations of our first approach became particularly apparent when it came to working with XML files of inscriptions rather than with databases. Jackson *et al.* discussed *LaQuAT* in more detail and also presents the results of our user acceptance testing.[13] The next section presents an approach that addresses these issues.

*5. Virtualization case study 2*

The JISC-funded Virtual Research Environment (VRE) project *gMan* investigated how to build a research environment for everyday data-driven research in Classics.[14] Specifically, it showed how to provide a variety of integrated views over heterogeneous archives that correspond to specific research interests and reflect the actual day-to-day working practices of the researchers that work with the resources. These practices range from highly specialized semantic annotations using community standards such as EpiDoc, to the use of standard, online search tools such as integrated library catalogues and Google Scholar (see Blanke *et al.* for more details and the results of the user evaluation).[15] In the *gMan* experiment, we wanted to support those communities of humanities researchers that would like to work with a specialized set of digital collections but are not satisfied with standard search and retrieval tools. These researchers may want to search across resources based on looser criteria of relevance – for example by searching for all Roman legal texts in one resource containing information on punishments that are also mentioned in papyri from another resource – and where their needs are served neither by the sort of search functionality investigated in Section 4 – which is too highly structured – nor by such very general-purpose search tools as Google, which will fail to deliver the specific functionality required.

*gMan* addressed services that would enable more general-purpose Classics research activities, such as integrating and organizing the heterogeneous and often unstructured digital resources through advanced discovery facilities. We investigated how the UK and European research infrastructure can be exploited to support data-driven, collaborative research in Classics by using the gCube environment,[16] which was developed by the EU-funded *D4Science* project.[17] gCube allows virtual research communities to deploy VREs on demand by making use of the shared resources of the European research infrastructure, and provides services that match closely the sort of information organization and retrieval activities that we identified as being typical in humanities research. It enables this use by virtualizing these resources in full-text indexes, which can be interrogated using various standard search and browse tools. This way, content can be delivered to Classics researchers more effectively, independently of the location and implementation of that content, and with special facilities provided for customizing the retrieval, management, and manipulation of the content.

---

[12] T. Blanke, M. Hedges, and S. Dunn, 'E-science in the arts and humanities – from early experimentation to systematic investigation', in *E-SCIENCE '07: Proceedings of the Third IEEE International Conference on e-Science and Grid Computing* ed. by? (Washington, DC 2006) Pages; M. Nentwich, *Cyberscience. Research in the Age of the Internet* (Vienna 2003).

[13] M. Jackson, M. Antonioletti, T. Blanke, G. Bodard, M. Hedges, A. Hume, and S. Rajbhandari, 'Building bridges between islands of data — an investigation into distributed data management in the humanities', in *Proceedings of the Fifth IEEE International Conference on e-Science* (Washington, DC 2009) Pages.

[14] *gMan* project website: <http://gman.cerch.kcl.ac.uk>.

[15] T. Blanke, L. Candela, M. Hedges, M. Priddy, and F. Simeoni, 'Deploying general-purpose virtual research environments for humanities research', *Phil. Trans. R. Soc. A* 368, 1925 (2010) 3813-28.

In our experiment, researcher communities were able to ingest the three data resources described in Section 2 into the gCube environment, which involves mapping the resources and their metadata into the generic data model used by gCube. Researchers were supported in this task by an import service that provides standard workflows for importing data, workflows that can be customized by using a simple scripting language. The environment also allows a variety of text-based indexes to be created for the collections, thus generating a number of different views onto the collections.

Using the imported collections, researchers could then deploy specific VREs to work on specific research questions by combining the data resources to which their virtual organization has access with tools and services that support interaction with the underlying data. Various search and browse tools offer access to the collections using keywords or geo-locations as entry points. Finds can be brought together in so-called virtual collections, which assemble references to items in existing collections. These virtual collections and the items in them can in turn be shared among the group of researchers that come together in the VRE. Other tools and services include a report-writing tool, as well as several annotation services.

*6. A Classics data service*

Our final aim is to integrate disparate, heterogeneous data sources for Classics using virtualization technologies. Our current design, based on the experiences and issues outlined in the *LaQuAT* and *gMan* experiments, is outlined in figure 2. Using OGSA-DAI, we integrate database resources with connectors, which allow users to query multiple remote databases as if they were a single virtual database. Using gCube, we join together remote document collections using a single, joint index of the textual sources. Again, remote, heterogeneous datasets can be queried. The final aim would be a network of integrating servers, *e.g.* for disciplines in humanities, maintained by a trusted arts and humanities data service, similar to existing services such as the Archaeology Data Service[18] in the UK or the Dutch DANS.[19] Each of these is a member of the European

---

[16] L. Candela, D. Castelli, and P. Pagano, 'gCube: a service-oriented application framework on the grid', *ERCIM News* 72 (2008) 48–49.

[17] *D4Science* website: <http://www.d4science.eu>.

[18] Archaeology Data Service website: <http://archaeologydataservice.ac.uk>.

[19] DANS website: <http://www.dans.knaw.nl>.

*DARIAH* project,[20] a European infrastructure for digital arts and humanities, which will take up some of the ideas expressed here.

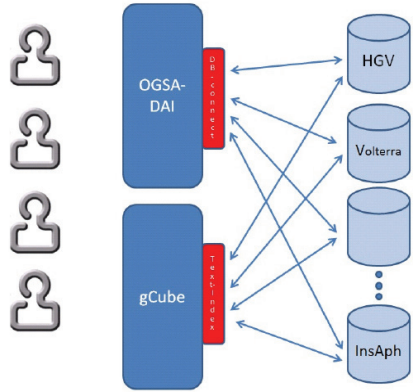

Figure 2: Classics virtual data centre

Thus, in our attempts to set up a virtual data centre for integrating Classics resources, to date we have two ways of connecting, developed by these two projects. Each way requires a trusted intermediary that would allow those remote sources to expose their data. This connection could be established in several ways – for example, the data could be transmitted and kept in a trusted vault, or the data centre could be allowed to query the remote data source directly. This will depend on the requirements of the remote data source and agreements made. Using gCube as an integration platform, the remote data source could also allow only the publication of the index of its textual resources.

*7. Conclusions and future work*

The main output for each case study was a demonstrator that provided integrated views over the three datasets used in the experiments (the same datasets were used in each case), although with quite different results in each. The conclusions from *LaQuAT* concerned limitations to the approach rather than solutions. The relational model followed by OGSA-DAI was more effective for resources that consist primarily of *structured* data (which we call *data-centric*) rather than for largely *unstructured* text (which we call *text-centric*), which makes up a significant component of the datasets we were using. This approach was, moreover, insufficiently flexible to deal with the semantic issues described in Section 2. The *gMan* project, on the other hand, addressed these problems by virtualizing data resources using full-text indexes, which can then be used to provide different views onto the collections and services that more closely match the sort of information organization and retrieval activities found in the humanities, in an environment that is more interactive, researcher-focused, and researcher-driven.

---

[20] *DARIAH* website: <http://www.dariah.eu>.

Subsequent to the projects described here, we have been funded to experiment with a linked data approach to the issues described as part of the *SPQR* project.[21] The primary aim of *SPQR* is to link and integrate datasets related to classical antiquity using RDF (Resource Description Framework) or equivalent formalisms, taking particular account to address the semantic issues described above. We will follow core standards, in particular the Europeana Data Model (EDM),[22] which has been developed by the EU-funded *Europeana* project for modelling cultural heritage data, as well as OAI-ORE (Open Archives Initiative Object Reuse and Exchange)[23] and emerging domain-specific ontologies and vocabularies. Ontologies form the centrepiece of the data integration project here, acting as semantic mediators for heterogeneous databases, which are mapped onto ontologies to provide semantic views over the datasets. *SPQR* is currently under development and we hope to report on results by the end of 2011.

It should be noted, however, that the resources used in the experiments described in this chapter were just three examples from among numerous others to which these various approaches could be applied. There are many small, scattered, yet related resources that would be much more useful to researchers if they were linked along these lines. Their utility would increase greatly once a certain critical mass is reached and together they would form a whole much greater than the sum of the parts, enabling researchers to ask questions that would not otherwise have been possible. An analogy might be a map, where each dataset represents a small area, say a few houses within a street; integrating a few of them is of limited utility, but after a certain point is reached there will be sufficient information to navigate from one place to another. A further output of the work is the definition of a Classics interoperability service for data resources, as defined in the previous section.

The benefits for the researchers include the ability to ask new questions by integrating the data resources. A great quantity of digital resources has been produced by humanities researchers in recent years, and a significant proportion of these are in the form of databases and of corpora of texts marked up in XML (usually some form of TEI). Although there are a number of initiatives to create standards in particular areas, such as EpiDoc, there will inevitably be a certain degree of variety in the representation of information gathered in different circumstances and for different purposes. A striking example is provided by museum databases, developed for the purposes of object cataloguing, and ill-suited to interact with other treatments of the objects within the museum. In any case, there is also a considerable body of legacy resources, especially databases, that exist in a variety of forms. However, there is a definite and pressing need to enable researchers to link up disparate data resources, but to do so using virtualization without affecting the original resources, which may be owned by different communities and subject to different rights.

Tobias Blanke (*King's College London*) tobias.blanke@kcl.ac.uk
Mark Hedges (*King's College London*) mark.hedges@kcl.ac.uk

---

[21] *SPQR* project website: <http://spqr.cerch.kcl.ac.uk>.

[22] EDM documentation: <http://pro.europeana.eu/edm-documentation>.

[23] OAI-ORE website: <http://www.openarchives.org/ore>.